\documentclass{aa}

\usepackage{graphicx}
\usepackage{txfonts}
\usepackage{hyperref}
\begin{document}

   \title{Optical-to-NIR magnitude measurements of the Starlink LEO Darksat satellite and effectiveness of the darkening treatment}
\titlerunning{Observations of Starlink's Darksat}

   \author{J. Tregloan-Reed\inst{1}
          \and             
          A. Otarola\inst{2,3}
          \and
          E. Unda-Sanzana\inst{4}
          \and
          B Haeussler\inst{3} 
          \and 
          F. Gaete\inst{3}
          \and
          J. P. Colque\inst{4}
          \and
          C. Gonz\'{a}lez-Fern\'{a}ndez\inst{5}
          \and
          J. Anais\inst{4}
          \and
          V. Molina\inst{4}
          \and
          R. Gonz\'{a}lez\inst{4}
          \and
          E. Ortiz\inst{6}
          \and
          S. Mieske\inst{3} 
          \and 
          S. Brillant\inst{3} 
          \and 
          J. P. Anderson\inst{3} 
          }
\authorrunning{Tregloan-Reed et al.
}

   \institute{Instituto de Investigación en Astronomia y Ciencias Planetarias, Universidad de Atacama, Copiapó, Atacama, Chile\\
              \email{jeremy.tregloan-reed@uda.cl}
              \and
              TMT International Observatory, 100 West Walnut Street, Pasadena, CA 91124, USA.
              \and
              European Southern Observatory, Alonso de Córdova 3107, Vitacura, Región Metropolitana, Chile 
              \and
              Centro de Astronom\'{i}a (CITEVA), Universidad de Antofagasta, Avenida U. de Antofagasta 02800, Antofagasta, Chile
              \and
              Institute of Astronomy, University of Cambridge, Madingley Road, Cambridge, CB3 0HA, UK
              \and
              Departamento de F\'{i}sica, Universidad de Antofagasta, Avenida Angamos 601, Antofagasta, Chile
             }

   \date{Received September 08, 2020; accepted January 06, 2021}

 
  \abstract
  {}
   {We aim to measure the Sloan {\it r'}, Sloan {\it i'}, {\it J,} and {\it Ks} magnitudes of Starlink's STARLINK-1130 (Darksat) and STARLINK-1113 low Earth orbit (LEO) communication satellites and determine the effectiveness of the Darksat darkening treatment from the optical to the near-infrared.}
   {Four observations of Starlink's LEO communication satellites, Darksat and STARLINK-1113, were conducted on two nights with two telescopes. The Chakana 0.6\,m telescope at the Ckoirama observatory (Chile) observed both satellites on 5\,Mar\,2020 (UTC) and 7\,Mar\,2020 (UTC) using a Sloan {\it r'} and Sloan {\it i'} filter, respectively. The ESO VISTA 4.1\,m telescope with the VIRCAM instrument observed both satellites on 5\,Mar\,2020 (UTC) and 7\,Mar\,2020 (UTC) in the NIR J-band and Ks-band, respectively.}
   {The calibration, image processing, and analysis of the Darksat images give {\it r}\,$\approx$\,5.6\,mag, {\it i}\,$\approx$\,5.0\,mag, J\,$\approx$\,4.2\,mag, and Ks\,$\approx$\,4.0\,mag when scaled to a range of 550\,km (airmass $=1$) and corrected for the solar incidence and observer phase angles. In comparison, the STARLINK-1113 images give {\it r}\,$\approx$\,4.9\,mag, {\it i}\,$\approx$\,4.4\,mag, J\,$\approx$\,3.8\,mag, and Ks\,$\approx$\,3.6\,mag when corrected for range, solar incidence, and observer phase angles. The data and results presented in this work show that the special darkening coating used by Starlink for Darksat has darkened the Sloan {\it r'} magnitude by 50\,\%, Sloan {\it i'} magnitude by 42\,\%, NIR J magnitude by 32\,\%, and NIR Ks magnitude by 28\,\%.}
   {The results show that both satellites increase in reflective brightness with increasing wavelength and that the effectiveness of the darkening treatment is reduced at longer wavelengths. This shows that the mitigation strategies being developed by Starlink and other LEO satellite operators need to take into account other wavelengths, not just the optical. This work highlights the continued importance of obtaining multi-wavelength observations of many different LEO satellites in order to characterise their reflective properties and to aid the community in developing impact simulations and developing mitigation tools.}


   \maketitle
%

\section{Introduction}\label{Sec:1}

In January 2020, Starlink launched its fourth fleet of 60 low Earth orbit (LEO) communication satellites. Due to major concerns from the amateur and professional astronomical communities (see \href{https://www.iau.org/news/announcements/detail/ann19035/}{IAU press release, 2019/06/03}, \href{https://www.iau.org/news/pressreleases/detail/iau2001/}{IAU press release, 2020/02/12}, \href{https://noirlab.edu/public/products/techdocs/techdoc003/}{NSF’s NOIRLab and AAS Technical Report}, and \href{https://noirlab.edu/public/products/techdocs/techdoc004/}{NSF’s NOIRLab and AAS Technical Report: Appendices}) along with Dark sky advocates (see \href{https://www.darksky.org/why-do-mega-constellations-matter-to-the-dark-sky-community/}{International Dark sky Association Press Release}\,; \citealt{Gallozzi2020}) regarding the reflective brightness of the LEO communication satellites, Starlink gave one of the LEO satellites launched in January 2020 a special darkening treatment. This satellite (STARLINK-1130), nicknamed `Darksat’, was the first attempt to reduce the reflective brightness of the satellites (SpaceX press kit, \href{https://www.spacex.com/sites/spacex/files/starlink_media_kit_jan2020.pdf}{January 2020}). The darkening treatment consisted of the white satellite surfaces being covered with a black diffuse applique,\footnote{Black diffuse appliques consist of a free-standing sheet of black material attached to a substrate.} while the phased array and parabolic antennas were painted with a specular black paint \citep[see][SpaceX updates \href{https://www.spacex.com/updates/starlink-update-04-28-2020/}{April 2020}]{Tyson2020}. 

Ground-based observations of Darksat show g\,$\approx$\,6.1\,mag \citep{Jeremy2020,Tyson2020} when normalised to the local zenith and accounting for the view angle subtended from the point of view of the satellite between the Sun and observer. When compared to STARLINK-1113, the darkening treatment used by Starlink reduced the reflective brightness of Darksat by $\approx$\,0.8\,mag \citep{Jeremy2020}, which is a reduction factor of two, while \citet{Tyson2020} determined an average reduction in the reflective brightness of Darksat by $\approx$\,1.0\,mag when compared to a group of Starlink satellites. The difference between \citet{Jeremy2020} and \citet{Tyson2020} highlights the complexity in accurately modelling the satellites and BRDF. \citet{Tyson2020} does not report a phase angle correction, only a range correction to the local zenith. When examining the range corrections from \citet{Jeremy2020} before the phase angle correction is applied, the magnitude difference between Darksat and STARLINK-1113 is $0.9\pm0.05$\,mag. When the statistical mean and standard deviation of the magnitude difference between Darksat and the non-darkened satellites from \citet{Tyson2020} is calculated, a result of $1.01\pm0.07$ is found, which is in agreement with \citet{Jeremy2020}, prior to the phase angle corrections.

The observed reduction in reflective brightness does not allow for non-linear image artefact correction on ultra-wide-field telescopes such as the Charles Simonyi telescope's camera detectors at the Vera C. Rubin Observatory. \citet{Tyson2020} shows that the Starlink LEO satellites need to be at least g $\approx$\,7.0\,mag to help ameliorate the impacts of electronic ghosts in the ultra-wide imaging exposures.

Several studies have been conducted to access the impact of LEO communication satellites on professional observatories. \citet{Hainaut2020} examined their impact on ESO telescopes in the optical and infrared and concluded that the greatest impact of LEO satellites will be on large telescopes using large fields of view (tens of arcmin to a few degrees) and relatively long exposure times (order of few seconds to minutes). The photometric model used by \citet{Hainaut2020} accurately predicts the magnitude of STARLINK-1113 given by \citet{Jeremy2020} when using the same satellite range and elevation. A second study by \citet{McDowell2020} indicates that LEO satellites with an orbital height of $\le600$\,km will have the greatest impact on twilight astronomy and long exposures using wide fields of view. The study also shows that satellites with a higher orbital height ($600$\,km to $2000$\,km), while being fainter, are visible longer through the night, with the greatest impact in summer. The satellites that are at an orbital height of 1000\,km, or above, will be illuminated by the Sun even at the local midnight. 

To date, all published observations (both professional and amateur) of the Starlink LEO communication satellites have been in the visible wavelength range. This is expected due to the concern raised by the professional and amateur astronomical communities and dark sky advocates regarding the naked eye visibility of the LEO satellites. However, there are some professional ground-based survey telescopes that observe at redder optical wavelengths (e.g. the Next Generation Transit Survey: \citealt{NGTS}; SDSS: \citealt{SDSS}) and in the near-infrared (NIR; e.g. VISTA: \citealt{VISTA_1,VISTA_2}). Properly gauging the impact that LEO satellites could have on these types of observations requires empirical magnitude measurements of the LEO satellites.

We present the first NIR observations ({\it J}-band: 1250\,nm; {\it Ks}-band: 2150\,nm) of Darksat (international designation: 2020-001U) and STARLINK-1113 (international designation: 2020-001N) LEO communication satellites along with observations in the Sloan {\it r'} (620.4\,nm) and Sloan {\it i'} (769.8\,nm) passbands. We then make comparisons of the reflective brightness between the two satellites for each passband and infer the wavelength-dependent effectiveness of the darkening treatment used on Darksat.

\section{Observations}\label{Sec:2}

STARLINK-1113 and STARLINK-1130 (Darksat) were observed by the Chakana 0.6\,m telescope on 5\,Mar\,2020 using a Sloan {\it r'} filter and on 7\,Mar\,2020 using a Sloan {\it i'} filter. Both Starlink LEO satellites were also observed using the ESO Visible and Infrared Survey Telescope for Astronomy (VISTA) telescope with the VISTA InfraRed CAMera (VIRCAM) on 5\,Mar\,2020 in the NIR J-band and on 7\,Mar\,2020 in the NIR Ks-band. In all cases, the observations were taken in twilight with the Sun between $-15^\circ$ and $-21^\circ$ below the horizon.

\subsection{Chakana 0.6\,m telescope}

The Chakana 0.6\,m telescope is based at the Ckoirama Observatory \citep{ckoirama}, the first Chilean state-owned observatory in northern Chile, and is managed and operated by the Centro de Astronom\'{i}a (CITEVA), Universidad de Antofagasta. The Chakana telescope (f/6.5) is equipped with an FLI ProLine 16801 camera, operated with one of three Sloan filters (g', r', and i'). In this setup, the CCD covers a field of view of $32.4\times 32.4$\,arcmin with a pixel scale of 0.47 arcsec pixel$^{-1}$. With the exception of the Sloan {\it r'} Darksat observation, which used no binning, the observations presented in this work from Ckoirama used $4\times4$ binning to reduce the CCD readout dead-time. 

By retrieving the Starlink two-line element (TLE) supplemental data from the Celestrak\footnote{\href{https://celestrak.com/NORAD/elements/supplemental/starlink.txt}{https://celestrak.com/NORAD/elements/supplemental/starlink.txt.}} website, our telemetry code\footnote{The telemetry code used in this work is written in Python and makes use of the \href{https://github.com/pytroll/pyorbital}{\tt pyorbital} package from the \href{https://github.com/pytroll}{\tt PyTroll project}. The telemetry code is available to the community by request from the author.} calculates the ephemerides (at a temporal resolution of 1\,s) of the satellites and the Sun using the longitude and latitude of the chosen observatory (Ckoirama geodetic GPS WGS\,84 coordinates: 24.0891\,S, 69.9306\,W, Altitude 966\,m). Table\,\ref{Tab.1} gives an observing log of the Chakana telescope observations presented in this work. The Chakana 0.6\,m telescope FITS files are available from the author by request.

\begin{figure*} \includegraphics[width=1.0\textwidth,angle=0]{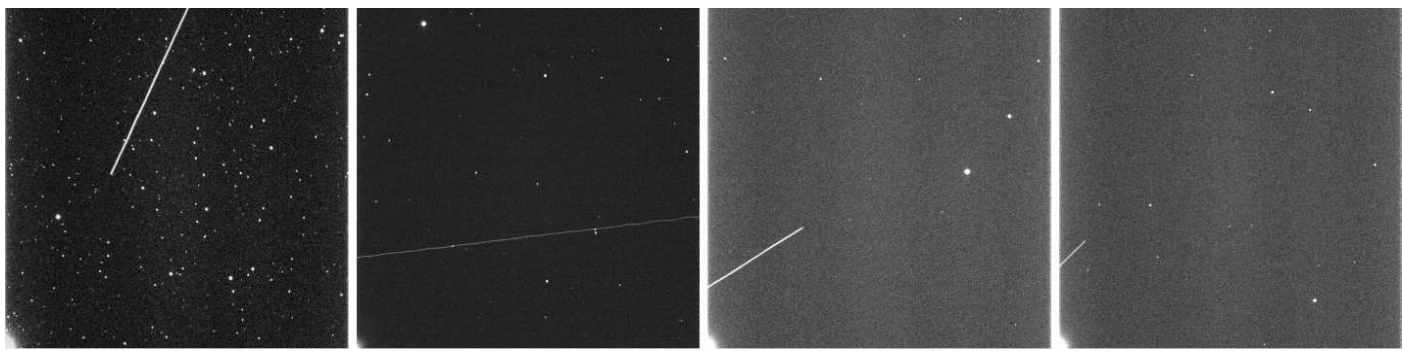} 
\caption{\label{Fig:1}FITS images of the satellite trails of STARLINK-1030 (Darksat) and STARLINK-1113 observed from Ckoirama. {\it left\textit{\emph{\textit{:}}}} STARLINK-1113: 5\,Mar\,2020 observed in Sloan {\it r'}. {\it middle left:} Darksat: 5\,Mar\,2020 observed in Sloan {\it r'}. {\it middle right:} STARLINK-1113: 7\,Mar\,2020 observed in Sloan {\it i'}. {\it right:} Darksat: 7\,Mar\,2020 observed in Sloan {\it i'}. } \end{figure*}

Figure\,\ref{Fig:1} shows the raw Flexible Image Transport System (FITS) images of the satellite trails of Darksat and STARLINK-1113, observed using the Chakana 0.6\,m telescope. Due to physical pointing restrictions at the Chakana telescope, it was only possible to capture a small section of the satellite trails on 7\,Mar\,2020.

\begin{table*} \centering
\caption{\label{Tab.1} Log of observations presented for STARLINK-1130 (Darksat) and STARLINK-1113.}
\setlength{\tabcolsep}{3pt} \vspace{-5pt} \begin{small}
\begin{tabular}{l|cccc|cccc} 
\hline\hline
 & \multicolumn{4}{c|}{STARLINK-1130} & \multicolumn{4}{c}{STARLINK-1113} \\
 & \multicolumn{4}{c|}{(Darksat)} \\
 \hline
 Date (J2000) & 5\,Mar\,2020 & 7\,Mar\,2020 & 5\,Mar\,2020 & 7\,Mar\,2020 & 5\,Mar\,2020 & 7\,Mar\,2020 & 5\,Mar\,2020 & 7\,Mar\,2020\\
 
 Time (UTC) & 00:34:27 & 00:27:01 & 00:29:55 & 00:23:58 & 00:20:37 & 00:11:55 & 00:14:54 & 00:09:07\\
 
 Telescope & Chakana & Chakana & VISTA & VISTA & Chakana & Chakana & VISTA & VISTA\\
 
 Filter     & Sloan {\it r'}& Sloan {\it i'}& {\it J}-band & {\it Ks}-band & Sloan {\it r'}& Sloan {\it i'}& {\it J}-band & {\it Ks}-band \\
 
 Exposure time (s)& 10.0 & 1.0 & 30.0 & 10.0 & 2.0 & 1.0 & 30.0 & 10.0 \\
 
 Altitude (km)& 564.01 & 562.08 & 563.73 & 560.28 & 563.32 & 562.77 & 564.32 & 560.75\\
 
 Range (km)& 866.39 & 991.73 & 1063.91 & 1146.11 & 718.89 & 880.06 & 1004.76 & 885.43\\
 
 Azimuth ($^\circ$)& 210.46 & 265.703 & 235.56 & 303.54 & 158.76 & 245.01 & 216.18 & 305.37\\
 
 Elevation ($^\circ$)& 37.72 & 30.93 & 30.51 & 24.86 & 49.63 & 36.78 & 30.51 & 36.28\\
 
 Airmass & 1.63 & 1.95 & 1.98 & 2.39 & 1.30 & 1.67 & 1.97 & 1.70    \\
 
 RA (Sat)& 03 06 51.26 & 02 15 0.25 & 02 04 23.17 & 03 17 59.93 & 08 23 44.29 & 02 29 25.40 & 01 44 05.66 & 03 43 29.02 \\
 
 DEC (Sat)& -60 47 16.89 & -40 18 18.00 & -40 57 33.32 & +16 14 44.37 & -59 30 02.72 & -33 40 21.07 & -57 37 41.22 & +10 11 39.22\\
 
 Angular velocity (arcsec\,s$^{-1}$) &$1239\pm35$ & $1491\pm39$ & $1064\pm33$ & $1305\pm36$&  $2996\pm55$ & $1438\pm38$ & $962\pm31$ & $2113\pm46$\\
 
 RA (Sun)& 23 04 24.31 & 23 11 48.29 & 23 08 06.40 & 23 15 29.81 & 23 04 22.16 & 23 11 45.96 & 23 08 04.08 & 23 15 27.52\\
 
 DEC (Sun)& -05 56 41.05 & -05 10 14.32 & -05 33 30.56 & -04 46 53.85 & -05 56 54.49 & -05 10 29.02 & -05 33 45.13 & -04 47 08.36 \\
 
 Sun--zenith angle ($^\circ$)& 110.7 & 109.5 & 109.4 & 108.6 & 107.6 & 106.2 & 106.1 & 105.3\\
 
 Sun--satellite angle ($^\circ$)& 72.3 & 51.6 & 53.2 & 64.8 & 117.3 & 56.8 & 61.3 & 70.5\\
\hline \end{tabular} \end{small}
\end{table*}

\subsection{ESO VISTA telescope}

The 4.1\,m VISTA telescope (f/3.25) is situated at the ESO Paranal observatory, Chile. It is equipped with an NIR camera, VIRCAM with a 1.65$^\circ$ diameter field of view, with a pixel scale of 0.34 arcsec pixel$^{-1}$. The camera is operated with five broad-band filters (Z, Y, J, H, Ks) and three narrow-band filters (NB\,980, NB\,990, and 1.18 micron; see also \citealt{VISTA_1,VISTA_2}).

The camera is comprised of 16 detectors arranged in a $4\times4$ matrix, with each detector having a $11.6\times 11.6$\,arcmin field of view. The gaps between the detectors are 10.4\,arcmin (horizontal) and 4.9\,arcmin (vertical) giving a nominal field of view of $1.292^\circ\times1.017^\circ$ on the sky. However, a single pointing (called an exposure) provides only partial coverage of the field of view. A five-point jitter was performed to remove the bright NIR sky-background this creates a stack image from the five exposures. This process was repeated six times with a different offset to obtain continuous coverage of the entire field of view, which is then combined to create a tile image. Because LEO satellite trails pass though the full field of view within a single pointing, and therefore only appear in a single exposure, a single five-point jitter was scheduled for the observations to allow the removal of the bright NIR sky background from the exposures.

\begin{figure*} \includegraphics[width=1.0\textwidth,angle=0]{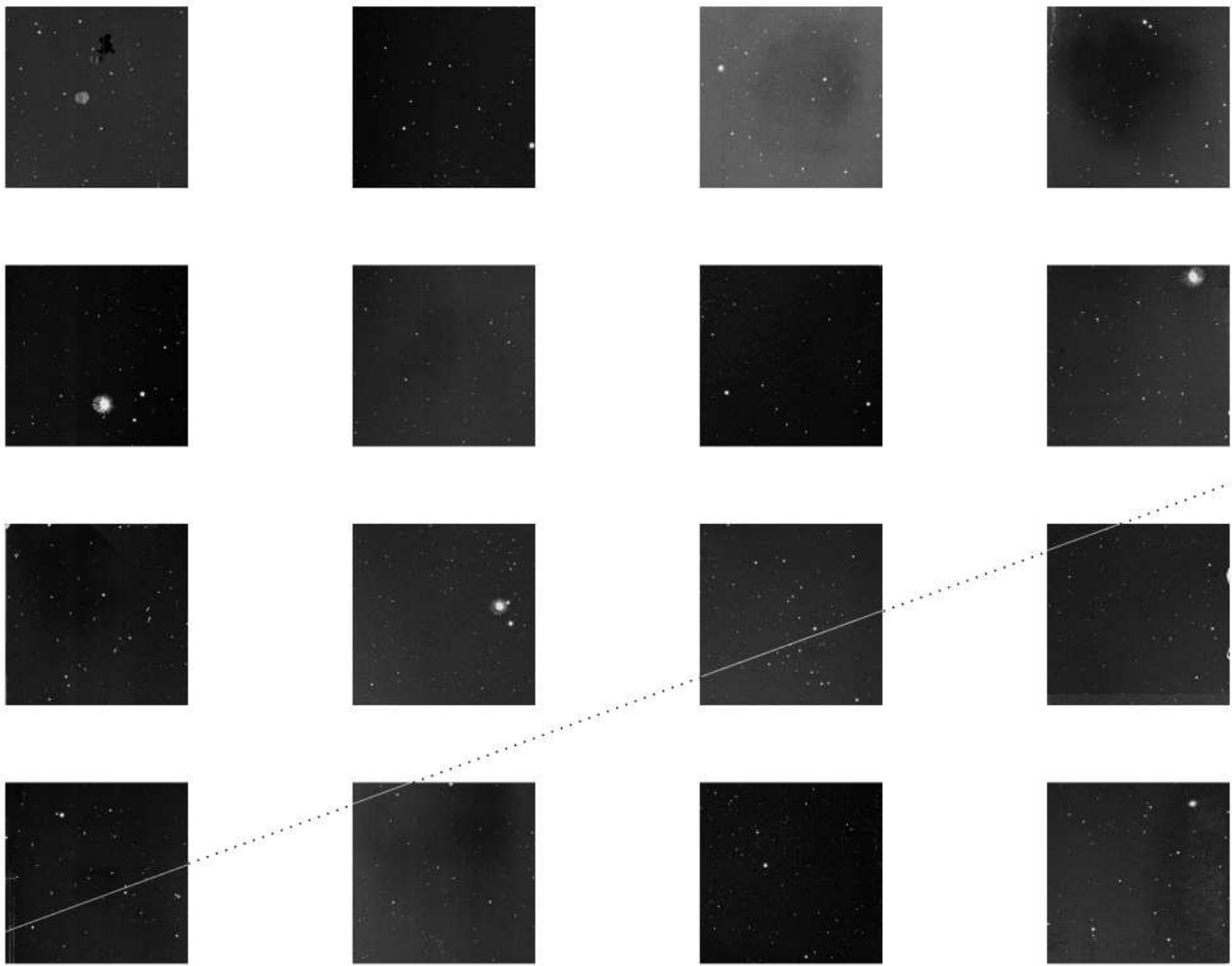} 
\caption{\label{Fig:2}{\it J}-band FITS images of the STARLINK-1030 (Darksat) satellite trail observed with the VIRCAM on 5\,Mar\,2020. The 16 detector images are arranged in the detector plane. Each detector covers an $11.6\times 11.6$\,arcmin field of view, and gaps between the detectors are scaled to 10.4\,arcmin (horizontal) and 4.9\,arcmin (vertical). The satellite trail that falls within the detector gaps is marked by a dotted line.} \end{figure*}

Figure.\,\ref{Fig:2} shows the FITS images from all 16 detectors of VIRCAM during the Darksat observation from 5\,Mar\,2020. The detector gaps are scaled and the Darksat satellite trail between the detectors is marked. The satellite trails for both Darksat (Fig.\,\ref{Fig:3}) and STARLINK-1113 (Fig.\,\ref{Fig:4}) observed on 5\,Mar\,2020 lie on four detectors. For the observations conducted on 7\,Mar\,2020, the satellite trails lie on three detectors for Darksat (Fig.\,\ref{Fig:5}) and four detectors for STARLINK-1113 (Fig.\,\ref{Fig:6}). An observing log of the VISTA telescope observations is given in Table\,\ref{Tab.1}. The VISTA data were obtained in twilight through technical run 60.A-9801(P), and they are publicly available at the ESO archives.\footnote{\href{http://archive.eso.org/cms.html}{http://archive.eso.org/cms.html}}
 
\subsection{Data reduction}

The FITS files from the Chakana 0.6\,m telescope were calibrated and reduced using the same methodology described by \citet{Jeremy2020}. The instrumental signature was removed from the images by subtracting the bias and dividing by the flat-field calibration frames. Standard aperture photometry was then conducted for each comparison star in the calibrated image.

\begin{figure} \includegraphics[width=0.48\textwidth,angle=0]{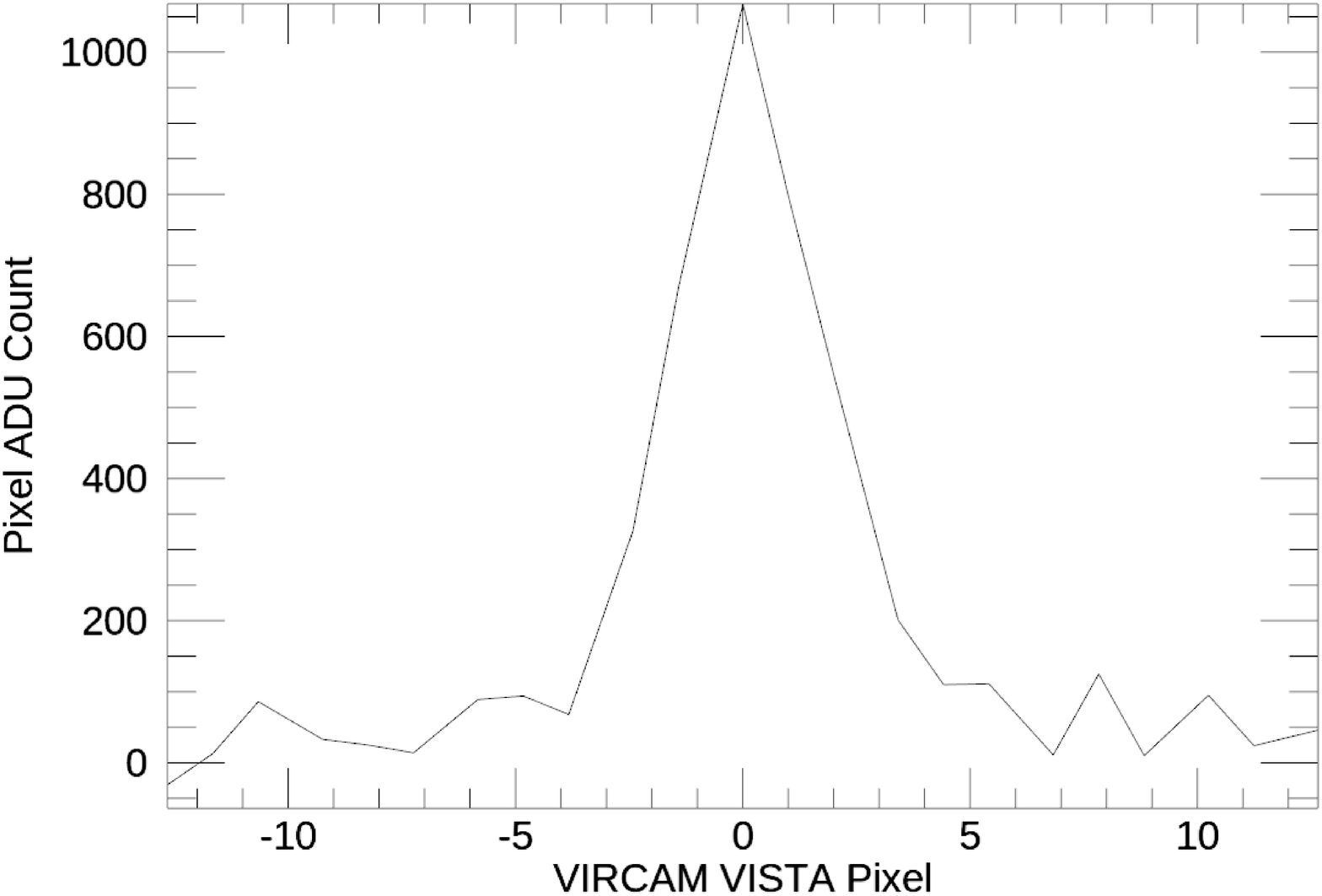} 
\caption{\label{Fig:2a} Transverse profile of a Darksat (STARLINK-1130) trail from detector 4 of the J-band VIRCAM VISTA observation on 5\,Mar\,2020. The FWHM was measured to be 4.5\,pixels (1.53\,arcsec) and shows a strong single peak in the transverse profile.} \end{figure}

It has been shown that there is a strong cut-off edge between the satellite trail and the sky-background for LEOsat trails observed on the 0.6\,m Chakana telescope \citep{Jeremy2020}. This is also true for for the images obtained using VIRCAM. Examination of the LEOsat trail transverse profiles in the VIRCAM J and Ks images show peaked point spread functions (PSFs), with a trail full width at half maximum (FWHM) = 4.5\,pixels or 1.53\,arcsec (Fig.\,\ref{Fig:2a}). For comparison, \citet{Tyson2020} simulated a LEOsat at an orbital height of 550\,km observed at a 50$^\circ$  elevation for the Rubin Observatory (8.4\,m telescope with f/1.234) and found the trail would be out of focus with a FWHM = 2.57 arcsec. This allows for two parallel lines to be fitted along the satellite trail edge and the total integrated flux of the satellite trail to be calculated, after subtraction of the sky background. Combined with the calibration from the comparison stars, the integrated flux of the satellite trail is converted into magnitudes.

The VIRCAM images are processed at the Cambridge Astronomy Survey Unit (CASU). A simple summary of the data reduction process is provided here; for more detailed information, we invite the reader to consult \citet{VISTA_Cal}. These images first undergo flat-field and gain correction, using a master flat-field frame obtained via a combination of several nightly sky flats. This master flat also allows for an initial gain correction to be applied by homogenising the measured sky level in all the detectors. By default, VISTA produces jittered images (sequences of pointings offset by a small, random amount), and these are combined with a recursive rejection algorithm to produce a sky frame. This sky emission is subtracted from all the individual images, along with the removal of readout artefacts (e.g. low-level pickup noise). An initial catalogue of detections is produced after this step, and cross-matching with 2MASS \citep{2mass} provides astrometric calibrators. A per-detector calibration is performed, with typical errors of 0.1\,arcsec.

These individual frames are normally sent downstream in the processing workflow to produce the standard data products, which are stacked (i.e. jitter corrected) and tiled (mosaics) images. However, these steps are designed in part to remove the signature of artificial satellites, and so for this study, individual calibrated exposures are used.

\section{Data analysis}\label{Sec:3}

For the Sloan {\it r'} and Sloan {\it i'} observations, the comparison stars were selected based on their respective signal-to-noise (S/N) above the sky-background and passband magnitude measurements present in the literature. For instance, while the Sloan {\it r'} Darksat observation on 5\,Mar\,2020 contains HD\,19801 in the field of view, the literature does not show a Sloan {\it r'} magnitude, and so HD\,19801 was not used as a comparison star. While it is possible to use a colour equation to convert the {\it B} and {\it V} magnitudes of HD\,19801 to a Sloan {\it r'} magnitude, with other comparison stars that have {\it B}, {\it V,} and r' magnitudes (e.g. TYC\,8860-962-1), we find that the colour equation $r' = V - 0.42(B-V) + 0.11$ \citep[Table\,1][]{Jester2005} does not agree with the measured Sloan {\it r'} measurements. Hence, to avoid unknown systematics, we avoided stars that did not have a Sloan {\it r'} or Sloan {\it i'} magnitude in the literature. Due to the short exposures and small aperture of the Chakana 0.6\,m telescope, we found only two to three comparison stars per image for the Sloan {\it r'} and Sloan {\it i'} observations. The comparison star magnitudes were selected from the following reference catalogues: the AAVSO Photometric All-Sky Survey DR\,9 \citep{AAVSOdr9} and RAVE DR\,4 \citep{RAVEDR4}. However, due to the longer exposure times and larger aperture of the VISTA telescope for the NIR observations, we used six to 12 comparison stars in each detector for the {\it J}-band observations, while four to eight comparison stars were used for the {\it Ks}-band detector images. The comparison star magnitudes were selected from the 2MASS \citep{2mass} reference catalogue.

\citet{Jeremy2020} discusses the fact that an estimated magnitude of an LEO satellite can be calculated when the satellite trail is not fully contained in an image. It is possible to extrapolate a predicted satellite trail length for a given exposure time from the angular velocity of the satellite (calculated from the satellite telemetry and given in Table\,\ref{Tab.1}). Then, by comparing the measured and predicted ephemerides trail lengths, an ephemerides-estimated magnitude is found.


For the VISTA observations, the satellite trails appear in multiple detectors. To simplify the analysis, each detector image was independently analysed. The integrated flux from each satellite trail was converted to magnitudes using only the comparison stars within the same detector image.

Table\,\ref{Tab.2} gives the weighted mean of the observed and estimated magnitudes from the different comparison stars along with the measured (length on detector) and predicted (multiplying the angular velocity of the satellite with the exposure time) ephemerides trail lengths for each detector, along with the time that the satellite spent on each detector. For the VISTA observations, each detector has a different observed satellite trail length. Therefore, we obtained an observed and estimated magnitude from each detector and then calculated the weighted mean for the estimated magnitude from the different detector results.

\begin{table*} \centering
\caption{\label{Tab.2} Results of the observations presented in this work for STARLINK-1130 (Darksat) and STARLINK-1113. The results are the weighted means of the individual results from each comparison star per detector. The weighted mean of the estimated magnitude after combining the different detector results is given in bold.}
\setlength{\tabcolsep}{8pt} \vspace{-5pt} 
\begin{tabular}{lccccccc} 
\hline\hline
Passband & Detector & Obs. trail length & Time on detector & Obs. & Est. trail length & Est.   \\   
          &          & (arcsec) & (s) &mag. & (arcsec)  &    mag. \\
\hline
\\
 \multicolumn{6}{c}{STARLINK-1130} \\
\hline
Sloan {\it r'} & 1 & $1938\pm44$ & $1.56\pm0.08$ & $9.29\pm0.01$ & $12386\pm111$ & $6.50\pm0.02$\\
Sloan {\it i'} & 1 & $202\pm14$ & $0.14\pm0.01$ &$8.50\pm0.02$ & $1491\pm39$ & $6.33\pm0.03$ \\
\hline
\\
\hline
{\it J}-band & 4 & $742\pm27$ & $0.70\pm0.05$&  $9.73\pm0.01$ & $31919\pm178$ & $5.64\pm0.02$  \\
{\it J}-band & 8 & $241\pm16$& $0.23\pm0.01$ &  $10.99\pm0.02$ & $31919\pm178$ & $5.68\pm0.02$ \\
{\it J}-band & 11 & $741\pm27$ & $0.70\pm0.05$&  $9.72\pm0.01$ & $31919\pm178$ & $5.65\pm0.01$ \\
{\it J}-band & 15 & $299\pm17$ & $0.28\pm0.03$&  $10.70\pm0.02$ & $31919\pm178$ & $5.63\pm0.02$ \\
\hline
 & & & & & & $\mathbf{5.65\pm0.01}$\\
 \hline
{\it Ks}-band & 11 & $735\pm27$ & $0.56\pm0.04$& $8.77\pm0.01$ & $13051\pm114$ & $5.65\pm0.03$\\
{\it Ks}-band & 13 & $332\pm18$ & $0.25\pm0.02$& $9.65\pm0.02$ & $13051\pm114$ & $5.66\pm0.03$\\
{\it Ks}-band & 14 & $410\pm20$ & $0.31\pm0.02$& $9.36\pm0.01$ & $13051\pm114$ & $5.60\pm0.02$\\
\hline
 & & & & & & $\mathbf{5.63\pm0.02}$\\
 \hline
 \\
 \multicolumn{6}{c}{STARLINK-1113} \\
\hline
Sloan {\it r'} & 1 & $1035\pm32$ & $0.35\pm0.02$& $7.37\pm0.03$ & $5993\pm77$ & $5.46\pm0.05$ \\
Sloan {\it i'} & 1 & $634\pm25$ & $0.47\pm0.03$& $6.32\pm0.03$ &$1438\pm38$ & $5.43\pm0.04$ \\
\hline
\\
\hline
{\it J}-band & 3 & $692\pm26$ & $0.73\pm0.05$&  $9.31\pm0.01$ & $28865\pm170$ & $5.11\pm0.01$ \\
{\it J}-band & 7 & $698\pm26$ & $0.73\pm0.05$&  $9.26\pm0.01$ & $28865\pm170$ & $5.08\pm0.01$\\
{\it J}-band & 11 & $698\pm26$& $0.73\pm0.05$ &  $9.27\pm0.01$ & $28865\pm170$ & $5.09\pm0.01$  \\
{\it J}-band & 15 & $698\pm26$ & $0.73\pm0.05$&  $9.31\pm0.01$ & $28865\pm170$ & $5.15\pm0.02$\\
\hline
 & & & & & & $\mathbf{5.10\pm0.01}$\\
 \hline
{\it Ks}-band & 8 & $254\pm16$ & $0.12\pm0.01$& $9.42\pm0.02$ & $21127\pm145$ & $4.62\pm0.04$\\
{\it Ks}-band & 11 & $775\pm28$ & $0.37\pm0.02$& $8.23\pm0.01$ & $21127\pm145$ & $4.63\pm0.03$\\
{\it Ks}-band & 13 & $400\pm20$& $0.19\pm0.01$ & $8.99\pm0.02$ & $21127\pm145$ & $4.68\pm0.03$\\
{\it Ks}-band & 14 & $524\pm23$ & $0.25\pm0.02$& $8.67\pm0.01$ & $21127\pm145$ & $4.66\pm0.03$\\
\hline
 & & & & & & $\mathbf{4.65\pm0.02}$\\
\hline
\end{tabular}
\end{table*}

When a Starlink LEOsat reaches its nominal orbit (on station), the solar arrays of the Starlink design are arranged above the main Starlink chassis. Reflected light from solar arrays is specular and is generally reflected towards space and not the Earth. When it is reflected towards the Earth, the glint lasts for a few seconds (e.g. satellite flares). However, since April 2020, Starlink -- in an effort to reduce the reflective brightness of the satellites -- has been automatically adjusting the solar array angles so that they are hidden behind the chassis, in order to prevent the solar arrays adding to the reflective specular brightness (see \href{https://www.spacex.com/updates/starlink-update-04-28-2020/}{SpaceX updates April 2020}).

The observed reflective brightness of an LEO satellite is dependent on the satellite range to the observer ($r$), the solar incidence angle ($\theta$), and the observer angle ($\phi$) of the satellite. The solar phase angle ($\alpha$) is the angle between the observer and the Sun as seen from the satellite. Therefore, to allow a direct comparison of the reflective brightness between Darksat and STARLINK-1113, the results presented in Table\,\ref{Tab.2} need to be corrected for $\theta$ and $\phi$ and normalised to a standard range. When observed at the local zenith (airmass = 1), $r$ is equal to the orbital height ($H_\mathrm{orb}$). We therefore normalise the magnitude of the satellites to the nominal $H_\mathrm{orb}$, 550\,km by scaling the magnitudes by $-5\log(r/550)$.

\begin{figure} \includegraphics[width=0.48\textwidth,angle=0]{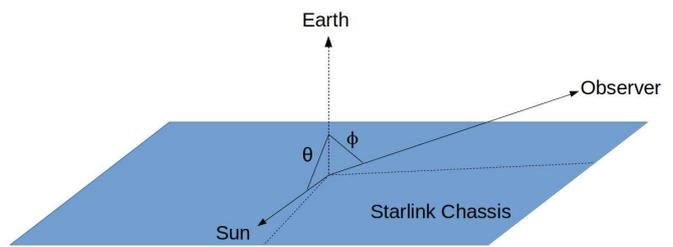} 
\caption{\label{Fig:2c}Diagram depicting the Earth-facing side of the Starlink chassis. The unit normal vector points towards Earth at the nadir. The vectors towards the Sun and observer are shown along with $\theta$ and $\phi$. } \end{figure}

The observer phase angle is the angle between the observer and the unit normal of the Earth facing surface of the satellite (see Fig.\,\ref{Fig:2c}). \citet{Jeremy2020} gave an equation to approximate $\phi$ using the straight line distance between the observer and the satellite footprint, nadir ($\eta$), elevation ($\epsilon$), and $H_\mathrm{orb}$:

\begin{equation}\label{Eq.1}
 \phi = \arcsin \left( \frac{\eta}{H_\mathrm{orb}} \sin\epsilon\right) \ .
\end{equation}

\begin{figure} \includegraphics[width=0.48\textwidth,angle=0]{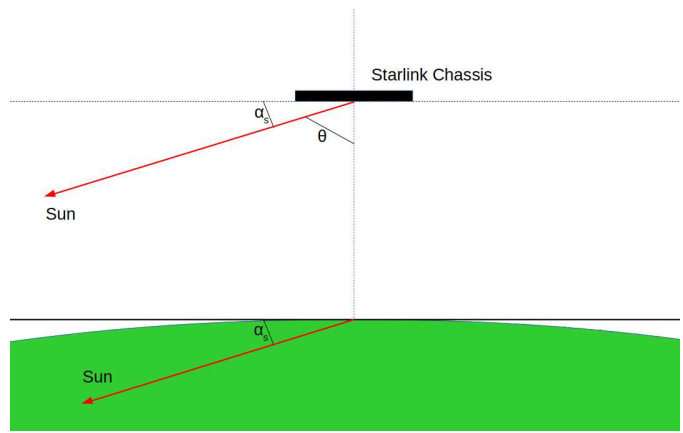} 
\caption{\label{Fig:2b}Diagram depicting the Starlink chassis parallel to the tangent at the nadir. The vectors from the Sun to both the satellite chassis and nadir are approximated as being parallel, and they form the angle, $\alpha_s,$ with the Starlink chassis and the tangent at the nadir, allowing $\theta + \alpha_s = 90^\circ$.} \end{figure}

The solar incidence angle (see Fig.\,\ref{Fig:2c}) can be calculated by evaluating the solar elevation angle ($\alpha_s$) at the satellite nadir. As shown in Fig.\,\ref{Fig:2b}, the body of the satellite is parallel to the tangent to the Earth's surface at the nadir. With the Sun at infinity, the two vectors from the Sun to the satellite chassis and from the Sun to the satellite nadir can be approximated as parallel. Therefore, $\theta + \alpha_s = 90^\circ$ when $\alpha_s$ is evaluated at the satellite nadir. The values of $\theta$, $\phi$, and $\alpha$ for all observations of Darksat and STARLINK-1113 are given in Table\,\ref{Tab.3}, and for $\theta$ they were determined using {\tt pyorbital} for the position (longitude and latitude) of the satellite at the time of the observations to find $\alpha_s$ at the nadir.

\begin{table*} \centering
\caption{\label{Tab.3} Estimated magnitudes of Darksat and STARLINK-1113 for each passband after correction for the solar incident and observer phase angles, then normalised to a range of 550\,km (one airmass), including the observed range and solar phase, incident, and observer phase angles.}
\setlength{\tabcolsep}{8pt} \vspace{-5pt} 
\begin{tabular}{lccccccccc} 
\hline\hline
Starlink&Obs.&Solar phase& Solar incidence& Observer &  Passband &Est. & Trail mag. \\  
satellite  &  range (km)& angle ($^\circ$)&angle ($^\circ$)&angle ($^\circ$)&  & scaled mag. & arcsec$^{-2}$\,s$^{-1}$ \\
\hline
1130 (Darksat)  & 866.39& 107.7 & 73.3 &45.1 &  r' & $5.63\pm0.07$ & $24.69\pm0.04$\\
1130 (Darksat)  & 991.73& 128.4 &77.4 &51.9 &  i' & $5.00\pm0.03$ & $24.03\pm0.02$ \\
1130 (Darksat)  & 1063.91& 126.8 &75.5&54.8 &  J & $4.21\pm0.01$ & $23.22\pm0.01$\\
1130 (Darksat)  & 1146.11& 115.2 &78.2 &57.7&  Ks & $3.97\pm0.02$ & $23.02\pm0.01$\\
\hline
1113 & 718.89 & 62.7 & 72.0& 35.9&r' &  $4.88\pm0.05$ & $23.87\pm0.04$ \\
1113 & 880.06 & 123.2 &79.3& 48.9&i' &  $4.41\pm0.04$& $23.36\pm0.03$ \\
1113 & 1004.76 & 118.7 &76.7 &51.8& J& $3.79\pm0.01$ & $22.76\pm0.01$\\
1113 & 885.43 & 109.5 & 81.4& 49.8& Ks &  $3.62\pm0.02$ & $22.53\pm0.01$\\
\hline \end{tabular} 
\end{table*}

Most of the observed reflective light from a complex body like a Starlink satellite is diffused. This effect can be approximated by using a bidirectional reflectance distribution function (BRDF). As the observations presented in this work are from a single point along the satellite trajectory path, we estimated the BRDF using a parametrised BRDF model from \citet{Minnaert1941} to give a first-order approximation:

\begin{equation}\label{Eq.2}
R = \left(\frac{\cos\theta_{1130}\cos\phi_{1130}}{\cos\theta_{1113}\cos\phi_{1113}}\right)^{k-1} \ ,
\end{equation}

\noindent where $R$ is the ratio of the solar phase attenuation between Darksat and STARLINK-1113; and $k$ is the Minnaert exponent and ranges from 0 to 1, with $k=1$ representing a perfect Lambertian surface. To approximate a dark surface, we set $k = 0.5$ \citep{brdf_book}.


In Table\,\ref{Tab.3}, we report the estimated magnitudes of Darksat and STARLINK-1113 from each passband, after correcting for $\theta$ and $\phi$ and then normalising to a range of 550\,km. The final results show that Darksat is dimmer than STARLINK-1113 by $0.75\pm0.06$ mag in the Sloan {\it r'} passband, $0.59\pm0.05$ mag in the Sloan {\it i'} passband, $0.42\pm0.01$ mag in the NIR J passband, and $0.35\pm0.02$ mag in the NIR K passband. 

To aid observers and observatories in determining the impact of LEOsat trails on the various detectors in the community, in Table\,\ref{Tab.3} we report the calculated trail brightness in units of mag\,arcsec$^{-2}$\,s$^{-1}$. The angular velocity of a satellite effects the amount of time that the satellite illuminates a pixel or arcsec$^{2}$ of a detector. Hence, after calculating the trail brightness in mag\,arcsec$^{-2}$, we used the angular velocity of each observation to determine the brightness as a function of time. This allows models of satellite trails to accurately estimate satellite trail brightness on different detector pixels.  

\begin{figure*} \includegraphics[width=1.0\textwidth,angle=0]{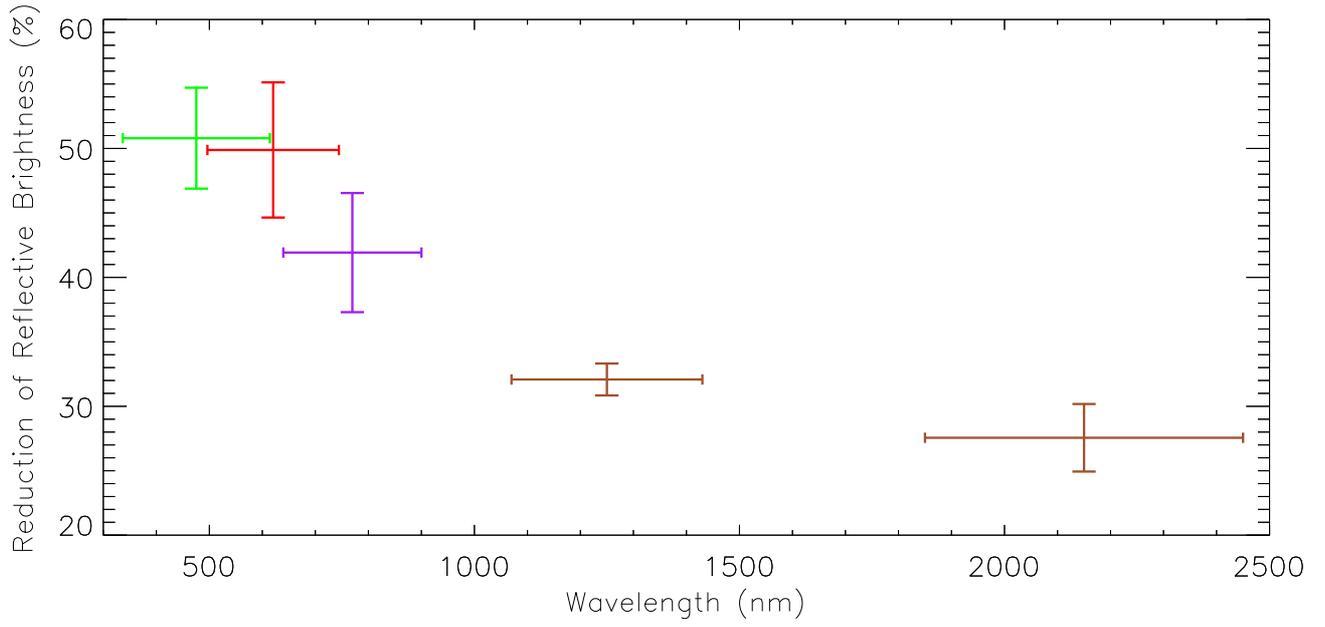} 
\caption{\label{Fig:7}Reduction of reflective brightness between Darksat and STARLINK-1113 for the results from this work: Sloan {\it r'} (red), Sloan {\it i'} (purple), J (brown), and Ks (brown); and from \citet{Jeremy2020}: Sloan {\it g'} (green). The horizontal error bars represent the FWHM of the passband filters, while the vertical error bars represent the uncertainty in the reduction.} \end{figure*}

\section{Summary and discussion}\label{Sec:4}

The Darksat and STARLINK-1113 results show that after correcting for the solar incidence and observer phase angles whilst normalising the range to the orbital height, 550\,km (one airmass), Darksat is dimmer than STARLINK-1113 in both the optical and NIR. However, the results also show that both satellites increase in reflective brightness with increasing wavelength (towards redder bands), and the effectiveness of the darkening treatment used for Darksat reduces with increasing wavelength. \citet{Jeremy2020} gave an estimated scaled magnitude of STARLINK-1113 and Darksat in the optical (Sloan {\it g'}) $5.33\pm0.05$ and $6.10\pm0.04$, respectively. Combining the results from this work with the results from \citet{Jeremy2020} shows that between 475.4\,nm (Sloan {\it g'}) and 2150\,nm (Ks), Darksat increases in reflective brightness by 2.13 mag ($\approx7$ times brighter). While STARLINK-1113 increases in reflective brightness by 1.71 mag ($\approx5$ times brighter).

The reduction in reflective brightness between Darksat and STARLINK-1113 is $\approx51\,\%$ (475.4\,nm), $\approx50\,\%$ (620.4\,nm), $\approx42\,\%$ (769.8\,nm), $\approx32\,\%$ (1250\,nm), and $\approx28\,\%$ (2150\,nm). Figure\,\ref{Fig:7} shows that the effectiveness of the darkening treatment used for Darksat is reduced from the optical to the NIR.

As discussed in \citet{Jeremy2020}, without BRDF measurements of Darksat and other non-darkened Starlink satellites, it is only possible to determine an estimated magnitude by using a parametrised BRDF model from \citet{Minnaert1941}. This approach takes the ratio of the solar phase attenuation between two satellites (Eq.\,\ref{Eq.2}) and removes the unknown BRDF component, which allows a comparison between the two satellites, irrespective of their attitudes at the time of the observations. However, by setting the Minnaert exponent to that of a dark surface, we obtain equivalent results to the first order approximation given by \citet{Hainaut2020} for the solar phase attenuation for a diffusing sphere, thus validating the two different approaches.

Performing future observations of Darksat and other Starlink satellites at multiple points along a single trajectory path will allow for an accurate BRDF measurement. The observations can then be compared to different empirical reflectance BRDF models (e.g. Phong BRDF: \citealt{Phong1975}; Lewis BRDF: \citealt{Lewis1994}) to improve the accuracy of the estimated magnitude measurements.

The Darksat satellite was Starlink's first mitigation experiment to reduce the reflective brightness of its LEO communication satellites. Simulations on the impact of mega-constellation LEO communication satellites show that the greatest impact to professional astronomy will be to ultra-wide imaging exposures from large telescopes \citep{Hainaut2020,McDowell2020}, such as the National Science Foundation’s Vera C. Rubin Observatory, formerly known as LSST. Studies on the mitigation of optical effects of bright LEO satellites on the Vera C. Rubin Observatory indicate that an optical magnitude of at least g\,$\approx7$\,mag is needed to allow for non-linear image artefact correction to the same level as background noise \citep{Tyson2020}. The results from this work and \citet{Jeremy2020} show that the reflective brightness of the standard Starlink satellite increases from 5\,mag in the optical to 3\,mag in the NIR, and Darksat increases from 6\,mag in the optical to 4\,mag in the NIR. This shows that the mitigation strategies being developed by Starlink and other LEO satellite operators need to take into account other wavelengths, not just the optical.

Since the launch of Darksat and subsequent observations \citep{Jeremy2020,Tyson2020}, Starlink launched its second mitigation experiment, STARLINK-1436, nicknamed `VisorSat', with its eighth launch in June 2020. VisorSat is equipped with deployable sunshades (transparent in radio frequencies), which act to block sunlight from the phased array and parabolic antennas. Apart from reducing the reflective brightness of the satellite, it is hoped that the sunshades will also prevent heating issues caused by the specular black paint used for Darksat \citep[see][SpaceX updates \href{https://www.spacex.com/updates/starlink-update-04-28-2020/}{April 2020}]{Tyson2020}. Future optical and NIR observations are planned for VisorSat and other Starlink non-darkened satellites using the Chakana telescope along with other mid-sized telescopes in Chile and Spain, to provide measurements of both the magnitude and the effectiveness of the VisorSat sunshades in the optical to the NIR.

\begin{acknowledgements}
      We would like to thank the anonymous referee for their helpful comments, which improved the quality of this manuscript. This work was supported by a CONICYT / FONDECYT Postdoctoral research grant, project number: 3180071. JTR thanks the Centro de Astronom\'{i}a (CITEVA), Universidad de Antofagasta for hosting the CONICYT / FONDECYT 2018 Postdoctoral research grant. Based on observations collected at the European Organisation for Astronomical Research in the Southern Hemisphere under ESO programme 60.A-9801(P). EU kindly acknowledges the work of Marco Rocchetto and Stephen Fossey to set up Ckoirama. The postgraduate students VM, JA, and RG of the Master of Astronomy programme of the Universidad de Antofagasta who are associated as co-authors of this manuscript acknowledges the Postgraduate School (Escuela de Postgrado) of the Universidad de Antofagasta for its support and allocated grants. JA and RG thank MINEDUC-UA project, code ANT 1795 for support. The postgraduate student EO of the Ph.D. programme in mathematical physics of the Universidad de Antofagasta who is associated as a co-author of this manuscript acknowledges the Postgraduate School (Escuela de Postgrado) of the Universidad de Antofagasta for its support and allocated grants. EO thanks the National Agency for Research and Development (ANID)/Scholarship Program / DOCTORADO BECAS NACIONAL CHILE/2018 - 21190387 for financial support. The following internet-based resources were used in the research for this paper: the NASA Astrophysics Data System; the ESO Online Digitized Sky Survey, the SIMBAD database and VizieR catalogue access tool operated at CDS, Strasbourg, France; and the ar$\chi$iv scientific paper preprint service operated by Cornell University.
\end{acknowledgements}

\bibliographystyle{aa}

\begin{appendix}
 
\section{VIRCAM VISTA FITS images}

\begin{figure*} \includegraphics[width=1.0\textwidth,angle=0]{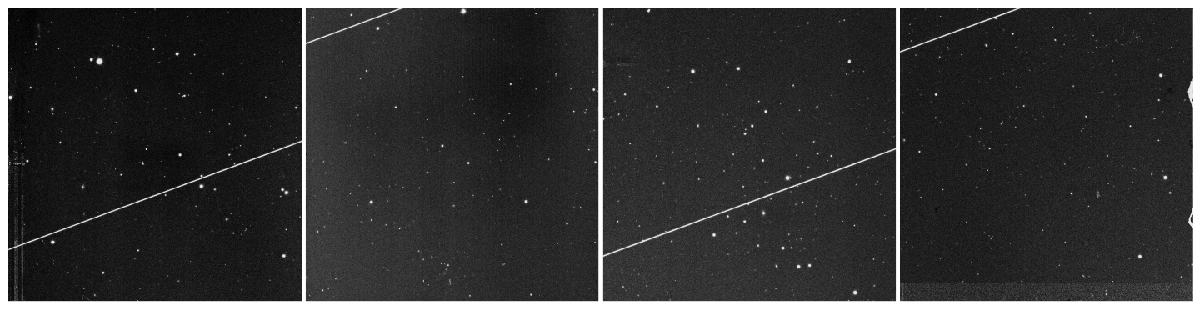} 
\caption{\label{Fig:3}Four detector FITS images of the STARLINK-1130 (Darksat) observed with VIRCAM on 5\,Mar\,2020 using the J-passband. The detectors are numbers 4, 8, 11, and 15.} \end{figure*}

\begin{figure*} \includegraphics[width=1.0\textwidth,angle=0]{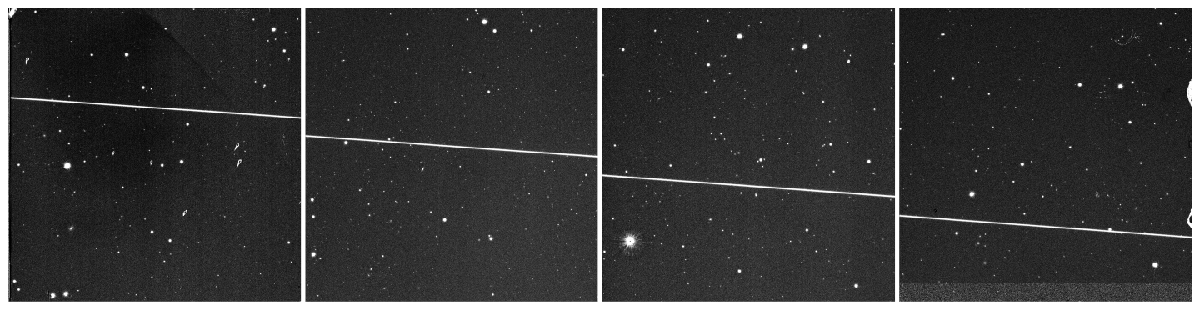} 
\caption{\label{Fig:4}Four detector FITS images of the STARLINK-1113 observed with VIRCAM on 5\,Mar\,2020 using the J-passband. The detectors are numbers 3, 7, 11, and 15.} \end{figure*}

\begin{figure*} \includegraphics[width=0.8\textwidth,angle=0]{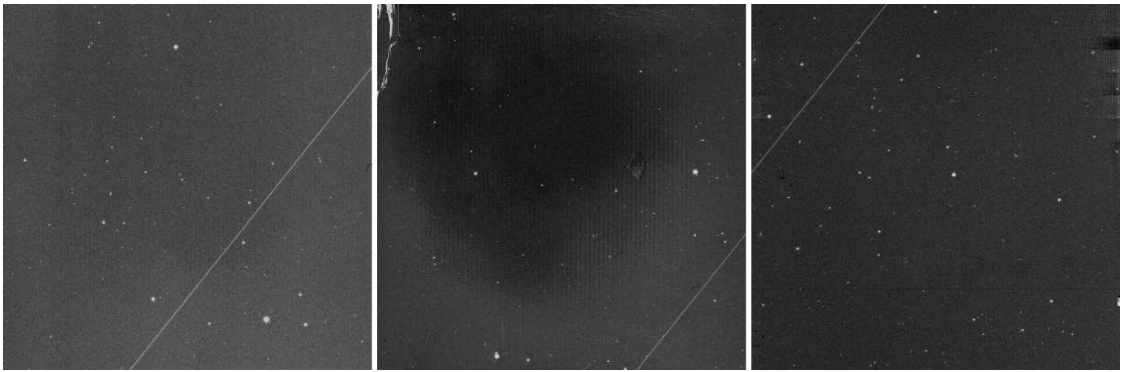} \centering
\caption{\label{Fig:5}Three detector FITS images of the STARLINK-1130 (Darksat) observed with VIRCAM on 7\,Mar\,2020 using the Ks-passband. The detectors are numbers 11, 13, and 14.} \end{figure*}

\begin{figure*} \includegraphics[width=1.0\textwidth,angle=0]{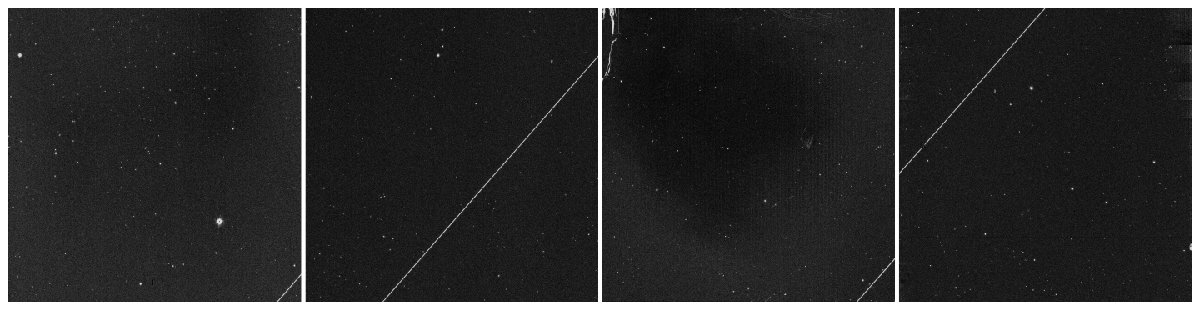} 
\caption{\label{Fig:6}Four detector FITS images of the STARLINK-1113 observed with VIRCAM on 5\,Mar\,2020 using the Ks-passband. The detectors are numbers 8, 11, 13, and 14.} \end{figure*}

\end{appendix}
\end{document}